# An Inverse Design Wavelength Demultiplexer for On-Chip Photoluminescence Sorting in TMDC Heterostructures

Anastasiia Zalogina[1,2,*], Chi Li[3,*], Ivan Zhigulin[1], Nathan Coste[1,2], Hossein Alijani[1,2], Otto Cranwell Schaeper[1,2], Hugo Charlton[1,2], Joseph Ward[3], Haoran Ren[3,*], Igor Aharonovich[1,2]

[1] School of Mathematical and Physical Sciences, University of Technology Sydney, Ultimo, New South Wales 2007, Australia

[2] ARC Centre of Excellence for Transformative Meta-Optical Systems, University of Technology Sydney, Ultimo, New South Wales 2007, Australia

[3] School of Physics and Astronomy, Faculty of Science, Monash University, Melbourne, Victoria 3800, Australia

* Corresponding authors: anastasiia.zalogina@uts.edu.au, chi.li1@monash.edu, haoran.ren@monash.edu

## Abstract

Emerging two-dimensional transition metal dichalcogenides (TMDCs) offer a promising platform for on-chip integrated photonics because of their unique optical and electronic properties. Their naturally passivated surfaces make them highly tolerant to lattice mismatch, enabling seamless heterogeneous integration by stacking different van der Waals materials, a crucial step in the development of advanced photonic devices. In this work, we present an inverse-designed wavelength demultiplexing waveguide capable of separating three distinct wavelengths. We further showcase its application in sorting and routing of distinct photoluminescence from the heterojunction formed by $WS_2$ and $WSe_2$ monolayers. The integrated nanophotonic chip splits and directs excitonic emission into individual waveguides at both room and cryogenic temperatures. Our demonstration opens new perspectives for integrating light sources in van der Waals materials with functional integrated photonics, offering a versatile platform for both fundamental research and practical applications.

Recent advances in integrating van der Waals (vdW) materials into photonic architectures have demonstrated their potential to enhance system scalability and performance [1-4]. These improvements stem from their novel optical properties and weak out-of-plane dangling bonds, which facilitate seamless integration onto chips. Compared to conventional semiconductors, vdW materials offer precise control over layer thickness and enable heterostructure stacking, making them ideal candidates for on-chip integration. A monolayer of transition metal dichalcogenides (TMDCs) is particularly attractive due to its exceptional electronic and optical properties [5-7]. Beyond their attractive electronic properties, TMDCs exhibit rich optical characteristics originating from excitonic photoluminescence (PL), which holds promise for photonic applications and provide a platform for exploring exciton physics relevant to spintronics and valleytronics [8-13].

Incorporating vdW materials into photonic devices allows the use of excitons for light-emitting applications, including examples of $MoTe_2$-based light-emitting diodes and photodetectors [14], field-effect transistors [15], and interlayer exciton emissions in $MoSe_2$/$WSe_2$ heterostructures for valley routing [16] and lasing [17]. Interlayer exciton emission is particularly interesting because it demonstrates significantly longer lifetimes than intralayer excitons, making it highly relevant for applications involving strong exciton correlations, and Bose–Einstein condensation [18-20]. Furthermore, recent advances in computational inverse design have begun to reshape the landscape of structures available to nanophotonics. Unlik0e traditional photonic design, which is usually based on intuitive rescaling or template reuse, inverse design employs optimization algorithms to discover optical structures based on desired functional characteristics with maximized performance [21-24].

Here, we present a wavelength demultiplexer based on an inverse design for on-chip sorting and guiding of distinct wavelengths. We further demonstrate its application in routing PL from stacked $WS_2$ and $WSe_2$ monolayers. Our nanophotonic chip is designed to efficiently sort and separate excitonic PL from $WS_2$, $WSe_2$, and interlayer excitons into three distinct waveguide channels. Figure 1 schematically illustrates the functional nanophotonic device. A heterostructure formed by stacking monolayers of $WS_2$ and $WSe_2$ functions as the light source of the device. This heterostructure has different band gaps and work functions of the two monolayers, resulting in a type-II band alignment. In this configuration, the conduction band minimum and valence band maximum are spatially separated in adjacent layers, allowing the formation of an interlayer exciton through the Coulomb interaction between electrons and holes confined to different monolayers [25-28]. These excitonic emissions are subsequently coupled into waveguide modes that are highly confined along the waveguide.

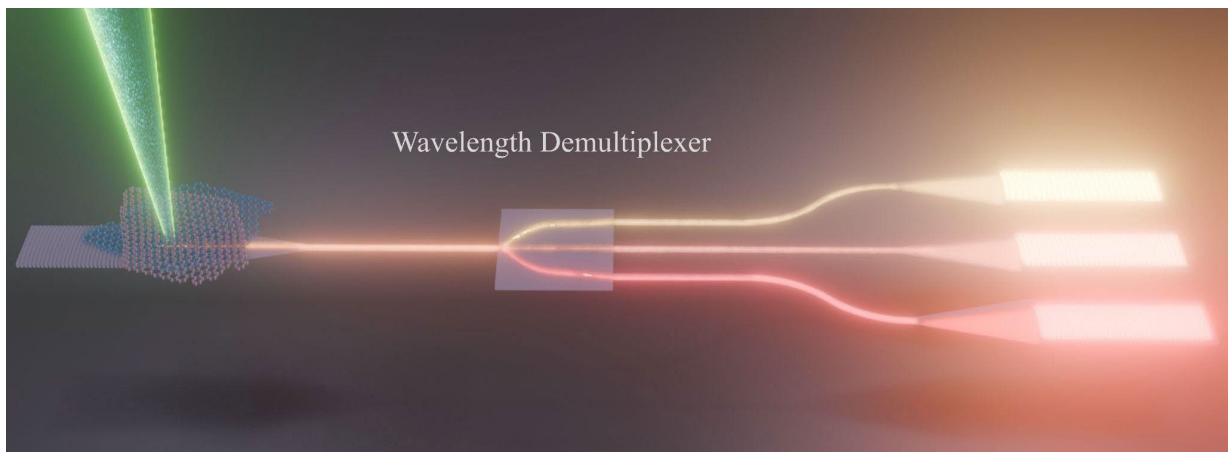

*Figure 1. Schematic of the nanophotonic chip based an inverse design wavelength demultiplexer for on-chip sorting of excitonic PL and its application for excitonic emission routing from TMDCs heterostructure. The heterostructure consists of $WS_2$ and $WSe_2$ monolayers stacked together and placed on a grating in-coupler. The distinct PL of the $WS_2$, $WSe_2$, and interlayer excitons is guided into the wavelength demultiplexer, which splits the emission into three outputs.*

To enhance the coupling efficiency and enhance far-field collection, we added grating-based in- and out-couplers at the ends that enable efficient incoupling of exciton PL into the waveguide, as well as outcoupling of the sorted emission to the far-field for optical detection. The exciton PL at three different wavelengths, at ~620 nm from the $WS_2$ exciton, ~750 nm from the $WSe_2$ exciton, and ~870 nm from the interlayer exciton, is guided to the inverse-design wavelength demultiplexer, which splits the PL into three separate outputs (see Figure 1). Our design is not limited to TMDCs but can be adapted for any optically active layered materials, offering a versatile platform for next-generation on-chip photonics.

We employ a gradient-based optimization approach for the demultiplexer design to sort the three excitonic emissions [22, 29-32]. As shown in Figure 2(a), the demultiplexer consists of low-loss silicon nitride ($Si_3N_4$) on top of a quartz substrate. To support broader spectral range, the waveguide dimensions (400-nm width and 314-nm height) of the input and output channels were selected to support the fundamental transverse electric (TE) mode.

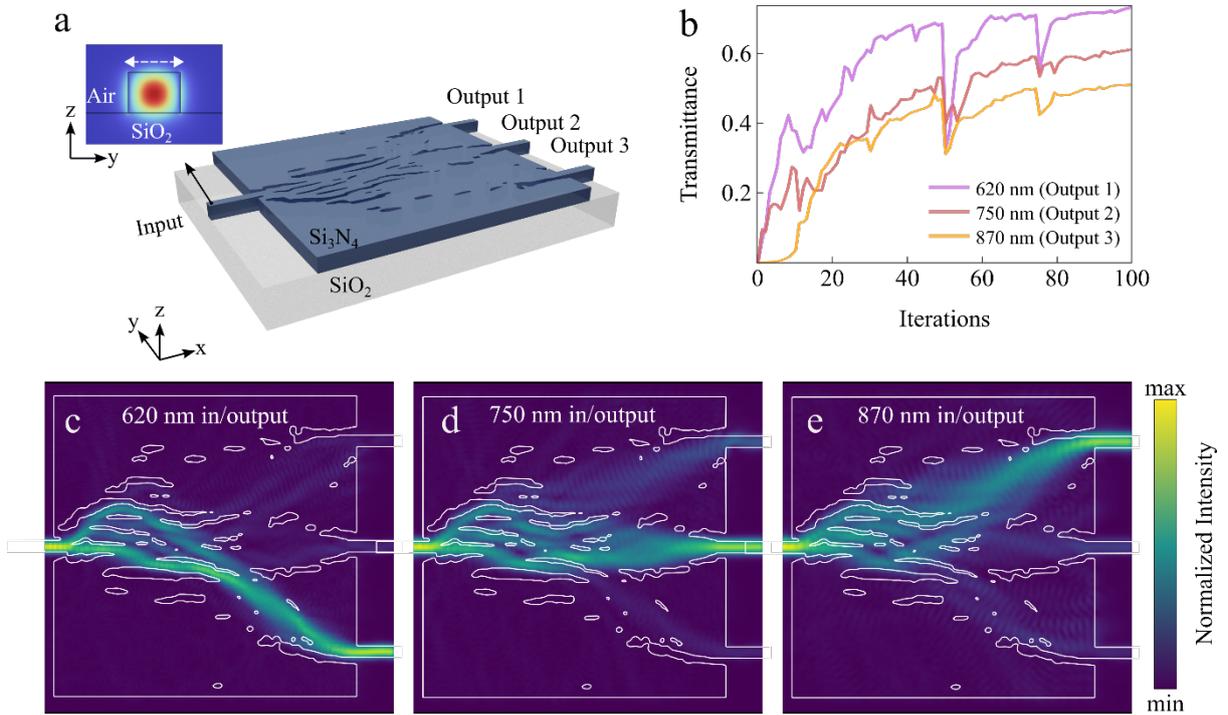

*Figure 2. Numerical simulations and design. (a) Demultiplexer configuration: a thin layer of $Si_3N_4$ on a quartz substrate with one input, a square demultiplexer region, and three patterned output channels. Inset: a fundamental TE waveguide mode confined in the input waveguide, the dashed arrow marks its polarization orientation. (b) Transmittance evolution for the three output channels. Electric field magnitude distribution of the demultiplexer at input of 620 nm (c), 750 nm (d), and 870 nm (e) after 100 iterations.*

An example TE mode profile is shown in the inset, where the electric field along the Y direction is seen to be largely confined within the $Si_3N_4$ structure. The corresponding effective mode indices for 620 nm, 750 nm and 870 nm are 1.76, 1.66 and 1.56 respectively. For comparison, fundamental transverse magnetic (TM) modes show a slightly lower effective index where the electric field aligns with the Z axis (see Figure S1 in Supplementary Information). To simplify, in the following optimization processes, we considered only the TE mode as the excitation source. Importantly, due to the small index difference, the TM mode was also confirmed to work for the optimized structure (see Figure S2 in the Supplementary Information).

The photonic problem starts with the optimization goal which is defined by an objective function, $f_{obj}$. The inverse-design wavelength demultiplexer uses an optimization algorithm to minimize this function for topology optimization. The process can be described as: $min_p f_{obj}(E(\epsilon(p)))$, where $p$ is a parameterization vector that is possible to practically fabricate any target device, $\epsilon$ is the permittivity distribution, E is the electric field distribution in three dimensions, and $f_{obj}$ is the objective function for global optimization. In short, optimizing the entire device performance means minimizing the objective function, which can be calculated based on an electromagnetic field simulation solver [22]. We define the objective function as $f_{obj}=(1-T_{620})^2+(1-T_{750})^2+(1-T_{870})^2$, where $T_{620}$, $T_{750}$, and $T_{870}$ are the transmittance at the corresponding output channels, which are shown in Figure 2(b) as the output transmittance versus the iteration number.

The inverse design program uses adjoint finite-difference frequency domain (FDFD) simulations to iteratively optimize the output powers of target modes, with the transmittance stabilizing after approximately 80 iterations, reaching final values of 74%, 60%, and 53% for the respective channels. To ensure fabrication compatibility, a minimum feature size of 60 nm was applied, and the optimization was validated using COMSOL simulations, showing consistent results despite slight transmittance variations due to binarization in the final discretization step (see more details in Section S1 of the Supplementary Information). Unlike conventional photonic design approaches, which often require extensive parameter sweeps, single-function devices based on inverse-design find non intuitive geometries that optimize performance and can enable access to new physical phenomena, offering deeper insights into light-matter interactions.

The device fabrication was done using electron beam lithography and pattern transfer with plasma etching (see details in Methods). Figure 3(a) shows a scanning electron microscope image (SEM) of the fabricated device.

To verify the functionality of the inverse-designed demultiplexer device, we first tested it using transmission measurements of pristine fabricated devices (that is, before transferring the heterostructure). Transmission measurements of the pristine fabricated device are shown in Figure 3(b-d). The device was characterized with laser light directed into the in-coupler at the designed wavelengths corresponding to the TMDC excitons and the interlayer exciton: 620 nm, 750 nm, and 870 nm (Figures 3 (b), (c), and (d), respectively). The collected signal from the out-couplers, which are output 1, output 2 and output 3, measured with an avalanche photodiode (APD), is presented and overlaid with SEM images for reference.

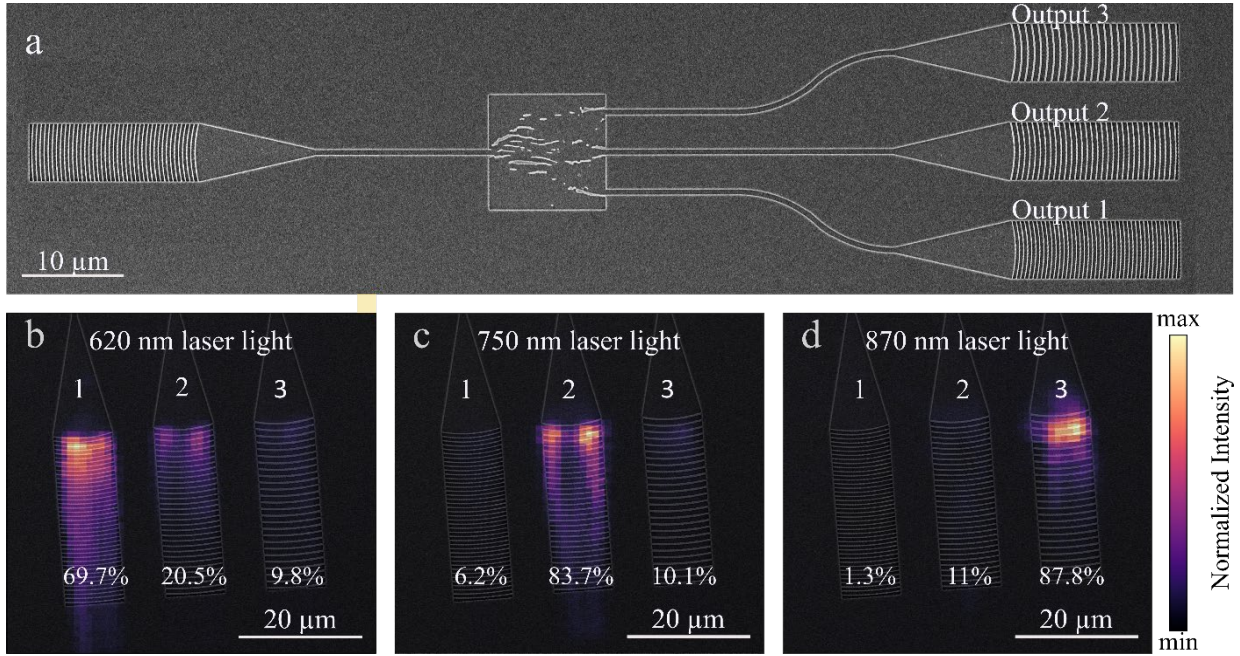

*Figure 3. Fabrication and functionality test of the device using laser light. (a) Top-view scanning electron microscope (SEM) image of the fabricated device. (a-c) Wavelength sorting with a laser before transferring the monolayers: normalized PL intensities detected using avalanche photodiodes, overlaid with SEM images for guidance. The device light guiding was tested with laser light at wavelengths of 620 nm (a), 750 nm (b) and 870 nm (c).*

As observed in Figure 3(b), most of the 620 nm light is guided into the designed output 1, although some light leaks into output 2. Figure S10 (b) in the Supplementary Information and Table 1 show that 69.7% of 620 nm emission is guided into output 1, 20.5% into output 2, and 9.8% into output 3. The 750-nm and 870-nm light were better sorted, as can be seen visually in Figure 3(c) and 3(d), respectively. From the 750-nm PL emission, 6.2% is guided into output 1, 83.7% into output 2, and 10.1% into output 3. And finally, for 870 nm, emission of 1.3% is guided to output 1, 11% to output 2, and 87.8% to output 3. These trends align with the data extracted from numerical simulations, which are overlaid with the experimental data in Table 1. In general, the device sorts the light according to the simulated design, including demultiplexing crosstalk that occurs between the channels. The primary sources of crosstalk are the final pattern binarization and difference in mesh resolution between the FDFD solver and COMSOL simulations, compounded by fabrication imperfections observed in the experimental measurements (see Section S2 in the Supplementary Information). The fabrication process introduced feature size limitations, where structures below 100 nm were not well resolved, likely contributing to increased crosstalk. This could be improved by optimizing electron beam lithography parameters, such as dose and proximity effect corrections, as well as refining the etching process to achieve sharper and more well-defined features. Additionally, non-polarized excitonic emission excites complex waveguide modes rather than a single fundamental mode, resulting in increased crosstalk. Incorporating multiple mode excitation into the inverse design could potentially mitigate this effect.

Table 1. Contrast fractions corresponding to experimental measurements and numerical simulations that were extracted from Figure 3(b-d) and Figure S2, respectively.

|  | Output 1 | | | Output 2 | | | Output 3 | | |
|---|---|---|---|---|---|---|---|---|---|
|  | Experiment | Theory (TE) | Theory (TM) | Experiment | Theory (TE) | Theory (TM) | Experiment | Theory (TE) | Theory (TM) |
| 620 nm | 69.7% | 95.9% | 84.6% | 20.5% | 0.9% | 0.7% | 9.8% | 3.3% | 8.4% |
| 750 nm | 6.2% | 20.8% | 2.6% | 83.7% | 74.5% | 92.8% | 10.1% | 4.7% | 4.6% |
| 870 nm | 1.3% | 2.9% | 6.9% | 11.0% | 37.5% | 12.0% | 87.8% | 59.6% | 81.2% |

After successfully testing the functionality of the inverse-designed demultiplexer device with laser light in Figure 3, we applied it to its designed goal, which is guiding exciton emission from a TMDC heterostructure. To prepare a $WS_2/WSe_2$ heterostructure, monolayers were mechanically exfoliated from bulk $WS_2$ and $WSe_2$ crystals, respectively. Exfoliating TMDCs down to monolayers allows their bandgaps to become direct for strong exciton emission [33, 34]. The monolayers were identified and selected using a combination of optical contrast and PL measurements (see Figure 4(b)), where peaks at 615 and 745 nm correspond to exciton emission in the $WS_2$ and $WSe_2$ monolayers, respectively. Using the dry transfer technique, the monolayers were transferred and stacked with a polydimethylsiloxane (PDMS) stamp. The $WS_2$ and $WSe_2$ monolayers are shown in pink and orange in Figure 4(a). The stack is encapsulated with hexagonal boron nitride (hBN) to protect it from the environment and prevent natural oxidation (dashed line in Figure 4(a)). Numerical simulations reveal that excess hBN partially covering the waveguide does not impair coupling efficiency, instead, it slightly enhances it by improving the overlap between the dipole and waveguide modes. The hBN-encapsulated excitonic emission couples efficiently to the waveguide in the near field, achieving an average coupling efficiency exceeding 20% (see Figure S3 in the Supplementary Information). While this efficiency is not exceptionally high, it surpasses the typical collection efficiency of high numerical aperture objective lenses in the far field (less than 10%). Further analysis of the demultiplexer regions reveals negligible disruption to the sorting capability, owing to the ultrathin hBN capping layer and its refractive index closely matching that of the $Si_3N_4$ waveguide (see Figure S4 in the Supplementary Information).

Achieving precise alignment with the device and placing the monolayers onto it is crucial. This was achieved from a home-built set-up that includes an optical microscope and micromanipulators (see Methods for details). TMDC flakes were transferred directly onto the surface of PDMS by mechanical exfoliation of the bulk material with tape. As the stamp and glass slide are transparent, the vdW flakes can be precisely aligned on the target substrate using the microscope and micromanipulators. Finally, the stamp is slowly peeled off to form a heterostructure.

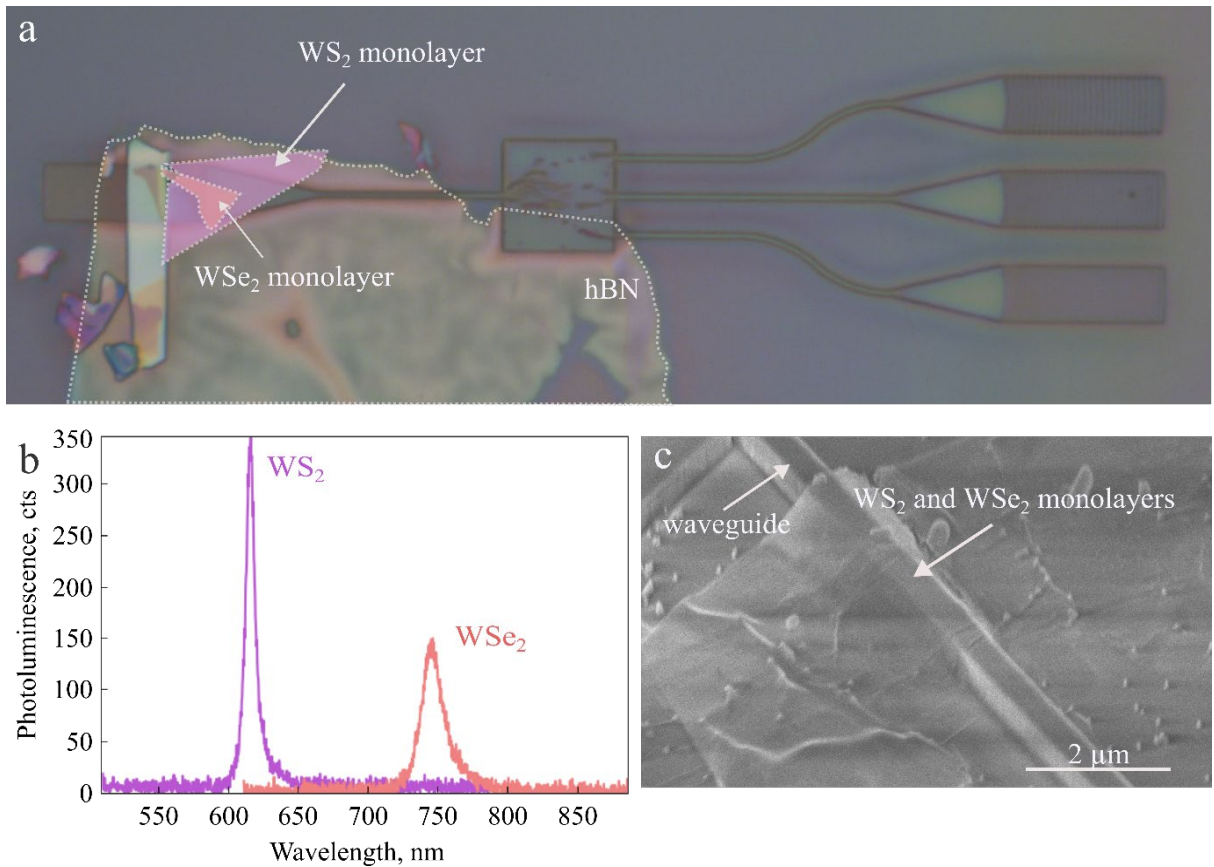

*Figure 4. TMDC heterostructure transfer. (a) Optical microscope image of WS$_2$/WSe$_2$ heterostructure on the fabricated device. The monolayers WS$_2$ and WSe$_2$ are indicated in pink and orange, respectively, while the hBN layer is shown in an uncolored dashed area. (b) Photoluminescence (PL) of the monolayers before their transfer onto the device. The peak wavelengths are 615 nm for WS$_2$ and 745 nm for WSe$_2$. (c) SEM image of the monolayers placed on top of the device.*

PL spectra were checked again after the transfer of WS$_2$/WSe$_2$ heterostructure (see Figure S10 (a) in Supplementary Information). The monolayers were excited with a 532 nm laser (from top, directly at the heterostructure), and the emission was collected through the out-couplers. Interlayer exciton PL at 850 nm was not observed at room temperature, likely due to its indirect nature in both real and momentum space, which typically requires cryogenic conditions for efficient radiative recombination [35].

Measurements at room temperature demonstrate overall guiding of exciton emission from WS$_2$ and WSe$_2$, consistent with the laser characterization results (Figure 5 (a, d-f)). The measured spectra from all three outputs (Figure 5 (d-f)) show that WS$_2$ exciton emission is guided into all three outputs. As seen in the PL map in Figure 5 (a), the light distribution along the gratings is uniform for the 620-nm output 1 and non-uniform for output 2. This distribution aligns well with our simulations (Figures S7 and S8 in the Supplementary Information), which confirms that the guiding of 620-nm emission into the middle output (Figure S7 (a) column 2), leads to significant scattering at the beginning of the coupler, which corresponds to the bright spots observed in Figure 5 (a). While the device may require further optimization for 620-nm wavelength coupling, WSe$_2$ exciton emission is effectively filtered into output 2 as shown in Figure 5 (e).

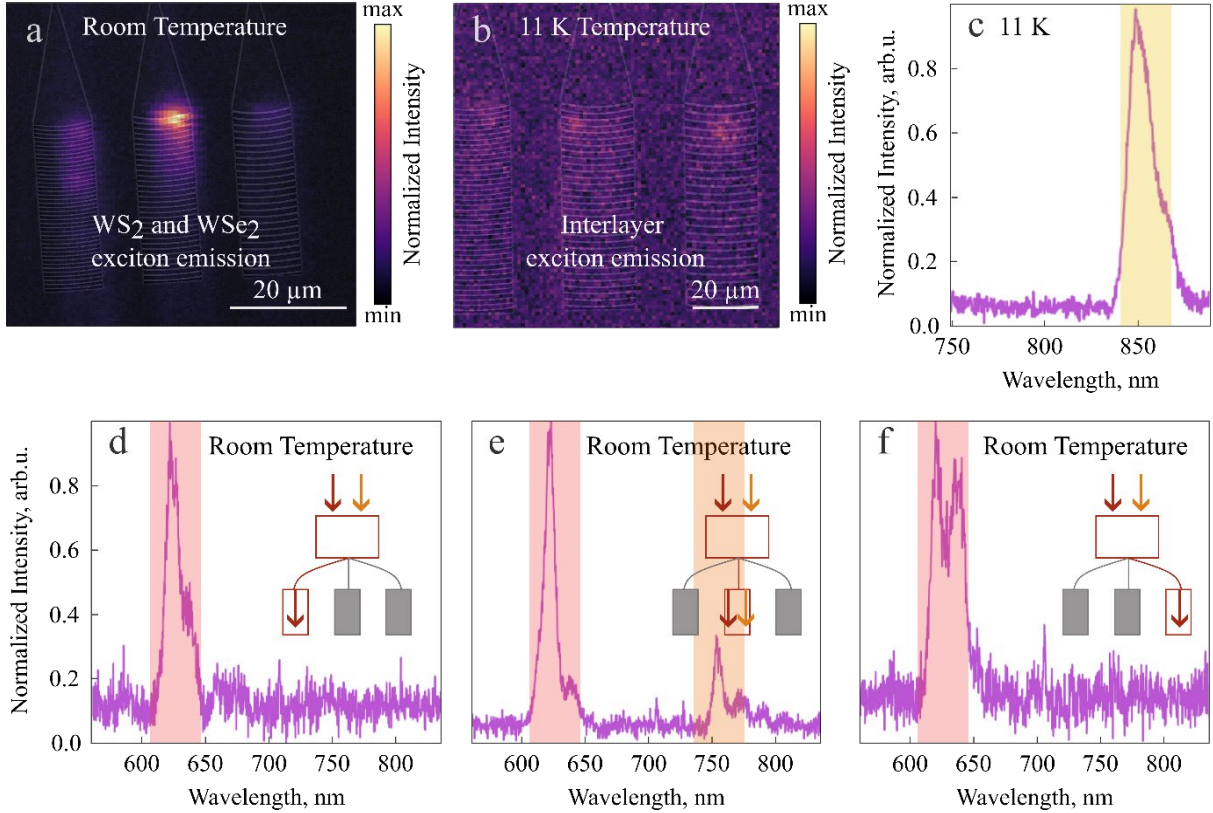

*Figure 5. Optical measurements after transferring the heterostructure, excited at a wavelength of 532 nm. (a, b) Normalized PL intensities overlaid with SEM images at room temperature (a) and 11 K (b). (c-f) PL spectra measured with the excitation at the in-coupler: at 11 K (c) and room temperature (d-f). The PL signal was collected at the in-coupler in (c), out-coupler 1 in (d), out-coupler 2 in (e), and out-coupler 3 in (f). The schematics of demultiplexer excitation and collection are shown in the insets of (d-f), where red arrows represent 620-nm exciton emission, and yellow arrows represent 750-nm exciton emission.*

Next, we measured the device with transferred heterostructure at 11 K where we observed interlayer exciton PL (see Figure 5(c)), confirming that while room-temperature emission was weak or suppressed, the excitonic transition remained accessible under cryogenic conditions. Interlayer exciton emission was successfully observed while collecting the signal from the in-coupler, with a strong peak at 850 nm (Figure 5 (c)). Here, WS$_2$ and WSe$_2$ exciton emission is absent, which can be explained by thermalization dynamics and charge transfer mechanism of interlayer exciton formation [36]. When two WS$_2$ and WSe$_2$ monolayers are stacked and optically excited, the holes from one monolayer tunnel to the other if the Coulomb interaction between them is strong enough. As a result, some or most of the electrons and holes participate in the formation of interlayer excitons, causing the PL intensity WS$_2$ and WSe$_2$ excitons to get less pronounced or disappear entirely. That explains the absence of a peak at 750 nm in Figure 5(c).

Guiding of interlayer exciton emission at the out-couplers was detected only using APDs (Figure 5b), as the signal was too weak for spectrometer-based measurements. The measurements show that the 850-nm emission is guided into all three channels, similar to the 620-nm emission, which also requires further optimization. The inverse design was based on the fundamental TE mode for simplicity, however, in practice, the guided modes that are excited by a dipole are hybrid combinations of TE and TM modes. For example, simulations for perpendicular excitation indicate that the 870-nm emission is distributed almost equally between the outputs 2 and 3 (48% and 48.3%, respectively), with only 3.7% directed to the output 1(see Table S2 in the Supplementary Information). The low intensity of the interlayer exciton emission further complicates experimental validation. These results confirm that interlayer excitons are indeed guided within the structure but remain difficult to detect with conventional spectroscopy, even at cryogenic temperatures.

We note that while the demultiplexing functionality was clearly demonstrated with laser excitation, the exciton emission results showed reduced sorting performance. This discrepancy can be attributed to lower and less directional coupling efficiency of broadband excitonic emission compared to the coherent laser sources used in the earlier test. These factors, combined with natural device-to-device variation, highlight the challenges in validating inverse-designed nanophotonic devices and should be taken into account in future optimization efforts and for integrating such a system into on-chip devices. While further optimization of coupling and loss performance is needed, inverse design remains a powerful tool for achieving multifunctionality in compact footprints. By enabling non-intuitive geometries that surpass the size and flexibility limits of conventional gratings and microresonators, inverse design supports scalable photonic integration beyond what traditional methods allow.

Realizing a fully integrated system with TMDC materials and inverse-designed photonics present several other practical challenges apart from the difficulties mentioned above. The transfer and alignment of the heterostructure onto the waveguide must be highly precise to avoid introducing strain or misalignment that could affect exciton coupling. Furthermore, scalability remains an open challenge, as variability in flake quality and interface cleanliness can affect reproducibility. Lastly, the sub-100 nm features required for the inverse-designed structure impose strict demands on nanofabrication accuracy.

Beyond TMDCs, the design is compatible with a wide range of optically active layered materials, including single-photon emitters such as defect centres in hBN or InSe. This adaptability opens opportunities for integrating our platform into emerging applications such as multiplexed quantum networks, where on-chip routing of spectrally distinct quantum emitters is essential [37-39].

In conclusion, we have demonstrated the advanced capabilities of an inverse-designed demultiplexer, achieving wavelength-selective routing of exciton PL at room temperature, that aligns closely with numerical simulations. Using this fabricated device, we successfully guided

WS$_2$, WSe$_2$, and interlayer exciton emissions into different channels. Although further optimization, particularly in exciton coupling efficiency, is still requires, our approach highlights the potential of integrating van der Waals materials with nanophotonic devices. This represents a significant step toward the realization of fully integrated van der Waals photonic systems and could enable scalable applications in quantum photonics.

**Methods**

**Numerical Simulations.** In the demultiplexer optimization, the waveguide width was set to 400 nm, while the demultiplexer was a 10-μm square region. A constant refractive index of 2.02 was used for all three wavelengths. Port/channel transmittances are calculated by square the mode overlap that is the ratio between output and input waveguide mode. The Python-based inverse design program was built on an open-source simulation framework named SPINS-B, shared by Stanford University. Electromagnetic field simulations were conducted using its built-in FDFD solver. To save computation time, we used a homogeneous mesh size of 50 nm for the 3D simulation. The optimization program was run on a high-performance laptop workstation equipped with an A1000 GPU and 20 GB of GPU memory. The optimized structure was further verified using COMSOL Multiphysics prior to nanofabrication.

**Fabrication.** To fabricate demultiplexers with waveguides, a 1000-nm SiO$_2$ layer was deposited with low-pressure chemical vapor deposition (Tystar Mini-Tytan), and a 314-nm layer of Si$_3$N$_4$ was deposited in the plasma chemical vapor deposition system (Oxford Instruments Plasmalab 100). Then, a layer of electron-beam resist poly(methyl methacrylate) (PMMA) A5 (AllResist gmbh) was spin-coated on the sample at 5000 rpm and 5000 acceleration for 30 s, followed by baking on a hotplate at 180° C for 3 min. The resist was then patterned by exposing it in the electron beam lithography setup (Elionix ELS-F125), with an area dose of electron beam exposure of 1400 μC/cm2, at 125 kV and 1 nA. The exposed pattern was developed in 1:3 IPA: MIBK for 60 s, rinsed in IPA for 20 s. The metal mask was then formed by depositing an 8-nm titanium and a 25-nm nickel using the electron beam evaporation system AJA ATC-1800-E and the subsequent lift-off process at 80 C using NMP (1-methyl-2-pyrrolidone). The designed pattern was transferred to the material using a reactive ion etching in the ICP-RIE system (Trion) at 8 mT, 12 sccm SF$_6$, 150 W RF, and 8 W ICP with the etch rate of ~2 nm/s. Finally, the metal mask is removed by wet etching using piranha solution, three parts of sulfuric (95-98%) acid H$_2$SO$_4$ and one part of hydrogen peroxide (30%) (H$_2$O$_2$) at room temperature. Electron microscope images of the resulting samples were obtained using a scanning electron microscope (Thermo Fisher Scientific Helios G4).

**Exfoliation and transfer.** The capping hBN flake, and WS$_2$ and WSe$_2$ monolayers were mechanically exfoliated with 3M Scotch tape on a polymer film (Gel-Fim® WF ×4, Gel-Pak, CA) [40]. The heterostructure stack of hBN/WS$_2$/WSe$_2$ was then assembled using a dry transfer method as follows. A polydimethylsiloxane (PDMS; SYLGARD™ 184 Silicone Elastomer, Dow, MI) stamp was prepared on a glass slide. Then a solution of polyvinyl alcohol (PVA), glycerol, and water (weight ratio 5:1:100) was drop-cast onto the stamp and dried on a hot plate at 42° C for 15 minutes. Next, the PVA/PDMS stamp was used to sequentially pick up the hBN, WS$_2$, and WSe$_2$ flakes at 70 °C, while aligning them with a custom-built setup. The completed stack was placed onto the target waveguide and released at 120° C.

**Optical measurements.** The fabricated device was tested with a supercontinuum light source NKT Fianium FIU-15 with a tunable VARIA frequency filter to guide light at 620 and 750 nm. The 870 nm channel was tested with a M-squared SOLSTIS Ti:Sapphire laser. Exciton emission was measured using a lab-built confocal microscopy setup with a fixed 532 CW laser excitation. The 4F system and scanning mirror allowed for decoupling of the collection from the excitation. A 100× 0.9 NA objective (Nikon) was used for both excitation and collection, directing light to a spectrometer (Princeton Instruments Acton SP2300) or an avalanche photodiode (APD) (Excelitas SPCM AQRH-14-FC) through a multimode fiber. Cryogenic optical characterization was carried out in a closed-loop helium cryostat for measurements at 12 K with a 100× 0.9 NA objective (Zeiss) for excitation and collection of light from the sample.

**Acknowledgements**

The authors acknowledge the use of the fabrication facilities and scientific and technical assistance from the Research and Prototype Foundry Core Research Facility at the University of Sydney, being a part of the NCRIS-enabled Australian National Fabrication Facility (ANFF), and the UTS and UNSW facilities, being a part of the ANFF-NSW node. The authors acknowledge Takashi Taniguchi (the National Institute for Materials Science) for providing hBN crystals and John Scott for silicon nitride substrates. This work was performed in part at the Melbourne Centre for Nanofabrication (MCN), the Victorian Node of the ANFF. The authors acknowledge financial support from the Australian Research Council (CE200100010, FT220100053), the Office of Naval Research Global (N62909-22-1-2028). H.R. acknowledges the funding support from the Australian Research Council (DE220101085, DP220102152).

**Supplementary Information**

The Supporting Information is available free of charge at

Extended version of numerical simulations (Section S1, Figures S1-S8), and additional data on fabrication (Section S2, Figure S9) and optical measurements (Section S3, Figure S10).

# Supplementary information

# An Inverse Design Wavelength Demultiplexer for On-Chip Photoluminescence Sorting in TMDC Heterostructures

Anastasiia Zalogina[1,2,*], Chi Li[3*], Ivan Zhigulin[1], Nathan Coste[1,2], Hossein Alijani[1,2], Otto Cranwell Schaeper[1,2], Hugo Charlton[1,2], Joseph Ward[3], Haoran Ren[3*], Igor Aharonovich[1,2]

[1] School of Mathematical and Physical Sciences, University of Technology Sydney, Ultimo, New South Wales 2007, Australia

[2] ARC Centre of Excellence for Transformative Meta-Optical Systems, University of Technology Sydney, Ultimo, New South Wales 2007, Australia

[3] School of Physics and Astronomy, Faculty of Science, Monash University, Melbourne, Victoria 3800, Australia

* Corresponding authors: anastasiia.zalogina@uts.edu.au, chi.li1@monash.edu, haoran.ren@monash.edu

**S1. Numerical Simulations.**

The inverse design program runs finite-difference frequency-domain (FDFD) adjoint simulations, running both forward and backward simulations in each iteration. Electromagnetic field gradients are calculated accordingly to monitor mode overlaps, which determines the objective function. The program iterates to maximize the output powers of the target modes in the corresponding channels. To ensure that the structure is amenable to standard nanofabrication procedures, a minimum feature size constraint of 60 nm was applied during the discretization step. Specifically, the optimization steps are divided into continuous optimization and discrete optimization. In the continuous optimization period, the material refractive index ($n$) varies continuously between air ($n = 1$) to $Si_3N_4$ ($n = 2.02$) while the discrete optimization applies a series of sigmoid functions to discrete the refractive index to make the pattern more discrete and finally to be binarized. In short, the continuous part allows us to explore more optimization possibilities, while discrete parts make the pattern fabricable with the price of a sudden efficiency drop. These can be seen in Figure 2 (b), where parameter adjustments for discretization lead to an abrupt drop in transmission.

To simplify the optimization processes, we considered only the fundamental TE mode for the excitation, as it is usually larger than the TM modes in a strip waveguide. For comparison, both the TE and TM waveguide modes are shown in Figure S1. The polarization for these two modes is perpendicular to each other, visualized by arrows. As a result of decreased field confinement, the mode index and intensity decrease with wavelength.

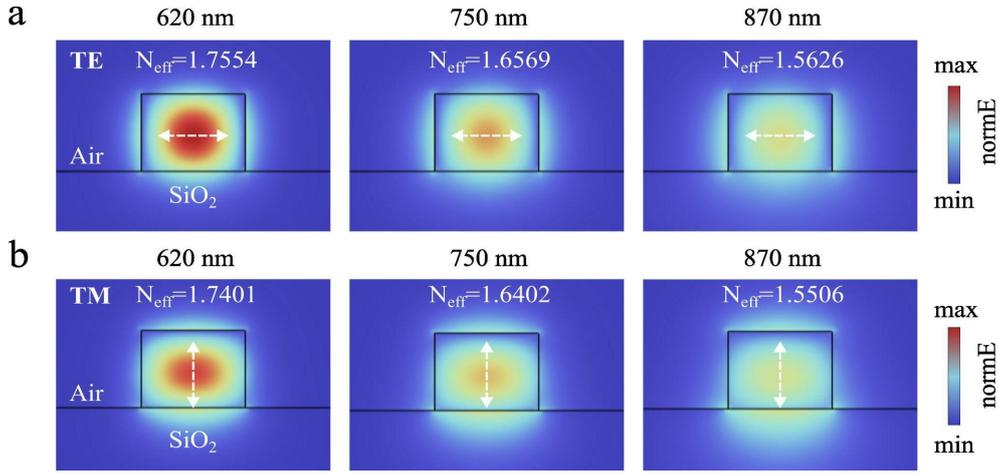

*Figure S1. Fundamental TE (a) and TM (b) waveguide modes at wavelength incidences of 620, 750 and 870 nm. Here, normE stands for electric field norm or magnitude, equals to $\sqrt{E_x^2 + E_y^2 + E_z^2}$. All colorbars set to the min and max intensities of the 620-nm TE mode. The dashed white arrows stand for the corresponding electric-field polarization.*

In our experiments, we added grating couplers to the waveguide to couple light to the TE mode, enabled by the polarization match between the laser and the gratings. However, the tightly focused beams as the excitation source could be more complex. Thus, although the design and optimization were performed in the TE mode, we verified the design's validity with both the fundamental TE and TM modes in COMSOL. The electric field magnitude distributions of the optimized demultiplexer, shown in Figure S2, demonstrate distinct wavelength-sorting as desired.

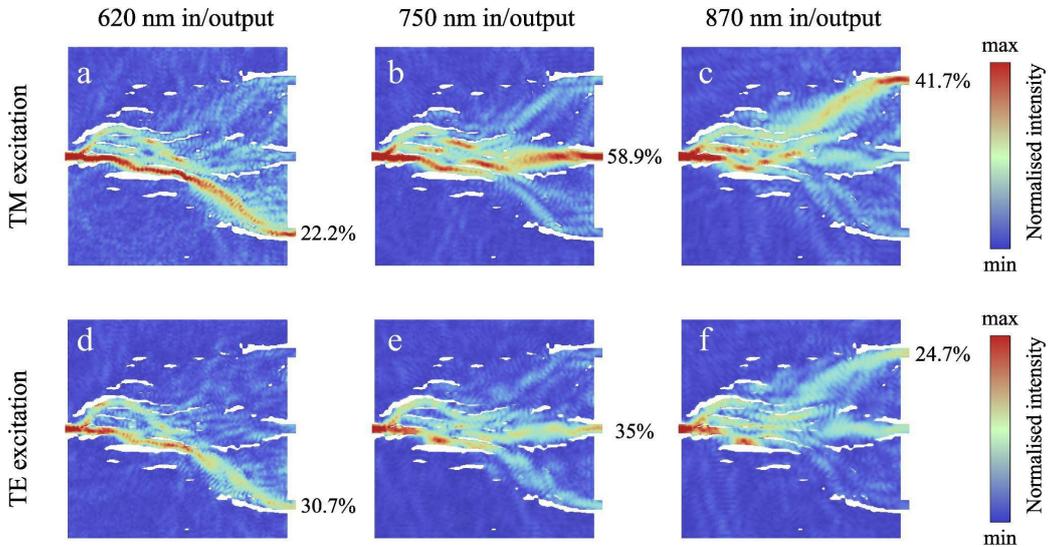

*Figure S2. Numerical simulation verification of the designed demultiplexer in COMSOL. Electric field magnitude distribution for fundamental TM mode inputs (a-c) and fundamental TE mode inputs (d-f) at 620 nm, 750 nm and 870 nm, respectively. Chanel transmittances are marked in the corresponding outputs.*

The transmittance varies slightly compared to the inverse design results. This is likely due to the binarization in the final step of the inverse design, where the optimization pattern is fully binarized to make the device fabricable. With more iterations, the transmission efficiency for all channels is expected to approach a balanced overall maximum governed by the objective function.

To estimate the coupling efficiency between excitonic dipole emission and the $Si_3N_4$ waveguide mode, we developed a numerical model of an electric dipole placed atop a waveguide and calculated it using COMSOL Multiphysics. The dipole was modelled as a point source with either parallel or perpendicular orientation to the waveguide propagation direction. The waveguide was either exposed or covered with a 20-nm thick hBN encapsulation layer, mimicking experimental conditions.

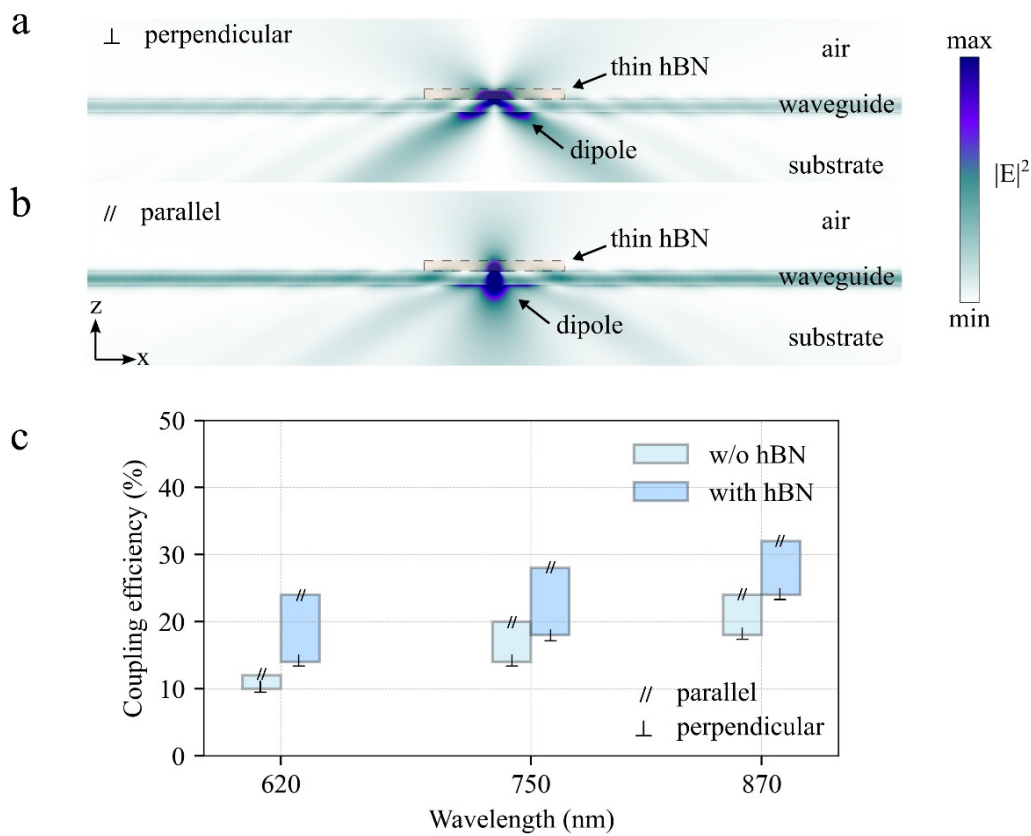

*Figure S3. An electric dipole coupled to a $Si_3N_4$ waveguide. Electric field power distribution for dipole moment perpendicular (a) and parallel (b) to the waveguide propagation direction. Dipole is capped under a 20-nm hBN flake. The example dipole emission wavelength is 750 nm in plotting (a) and (b). (c) Coupling efficiency dependency as a function of exciton emission wavelength and dipole orientation.*

The coupling efficiency was defined as the fraction of power emitted by the dipole that is coupled into the waveguide mode. This was quantified by calculating the time-averaged power flow through a spherical surface surrounding the dipole (representing total emission) and comparing it to the power propagating through a waveguide cross-section located 30 μm downstream. As shown in Figure S3(c), coupling efficiencies ranged from 14–32% for hBN-encapsulated dipoles and 10–24% without encapsulation, depending on dipole orientation and emission wavelength. The presence of hBN enhances coupling by increasing the effective mode volume of the waveguide, resulting in improved spatial overlap with the dipole's electric field. These values notably exceed the collection efficiency of a typical high-NA (0.9) microscope objective, which is usually below 10%.

Next, we investigated the potential impact of hBN coverage more closely. In our sample, hBN covers the waveguide region and partially extends into the demultiplexer region, as illustrated in Figure 4a. The presence of thin hBN slightly improves dipole-waveguide coupling (Figure S3 (c)), as the extended hBN increases the waveguide mode volume, better aligning it with the dipole's location. To further investigate the demultiplexer region in simulation, we covered the 620-nm channel with a thin hBN flake (same as the real sample) (Figures S4). Our results reveal negligible influence from the additional hBN on the final sorting performance, suggesting a robust design. This is likely attributable the close refractive indices of hBN (~2.1) and $Si_3N_4$ (~2.0) in the investigated range, along with the thinness of the capping layer, which is not substantial enough to significantly alter the light propagating within the demultiplexer. In conclusion, the coverage of thin hBN (< 20 nm) barely changes the demultiplexing performance.

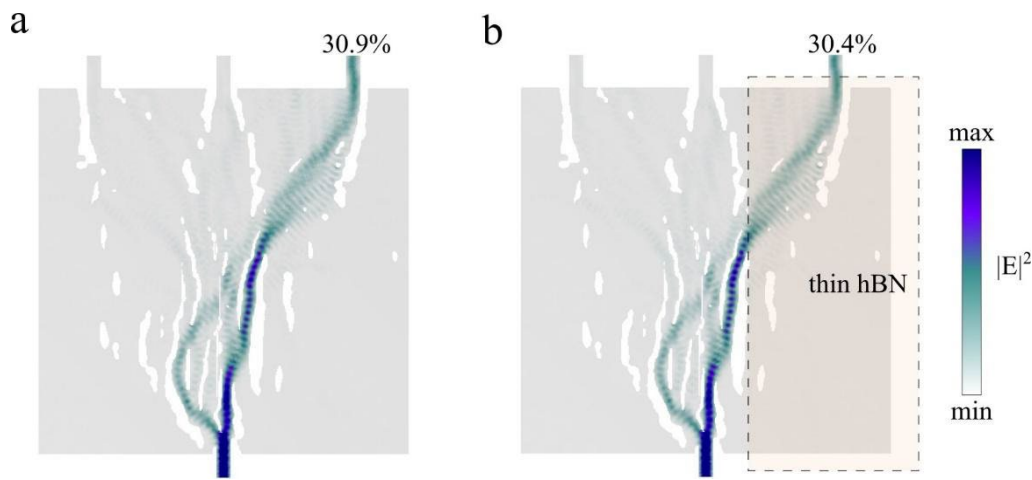

*Figure S4. Demultiplexing performance comparison with and without hBN coverage. (a) no hBN. (b) hBN partially covers the 620-nm channel. The input excitation is upward fundamental TE waveguide mode.*

To capture exciton emission coupling to the waveguide and demonstrate wavelength sorting in the demultiplexer, we placed an electric dipole on the waveguide's top surface (Figures S5 and S6). Perfectly matched layers were incorporated at both waveguides ends to ensure accurate power flow calculation, critical for evaluating the sorting performance of each output port.

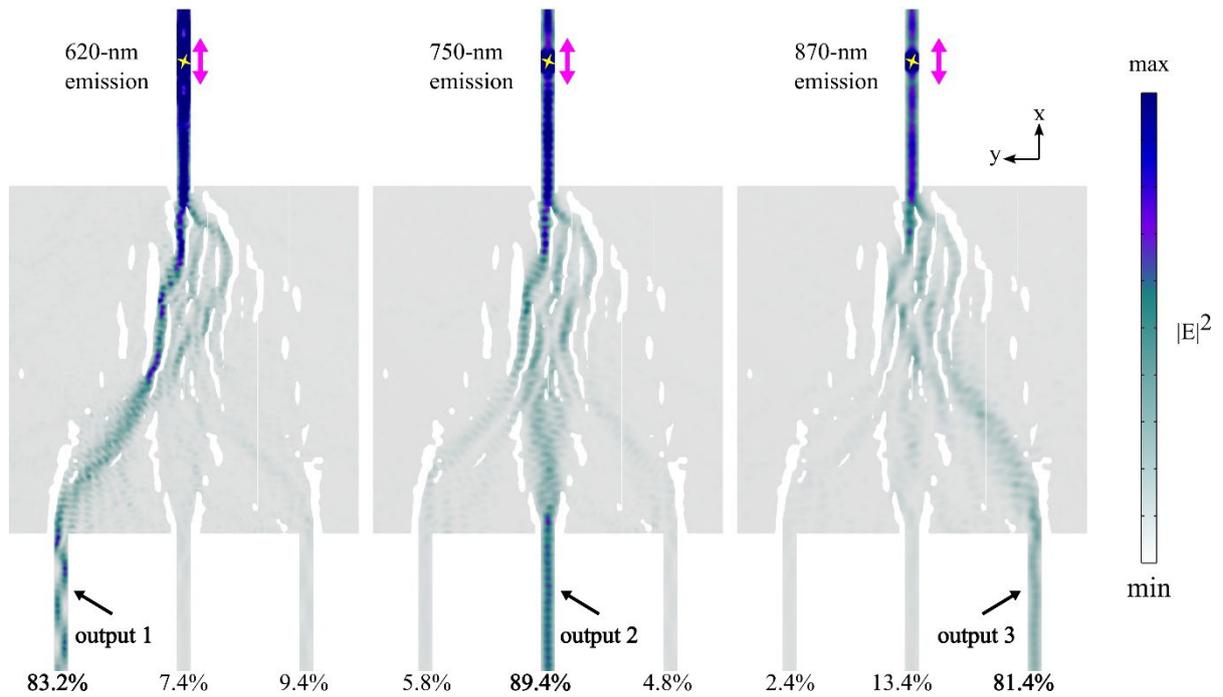

*Figure S5. Full device simulation with dipole source parallel to the waveguide. The values at each output port are transmission fractions. All plots normalized to the same value.*

We integrate time-averaged power flow across the input waveguide cross-section and the three output waveguide cross-sections. Transmittance was then defined as the fraction of output power to input power. The simulation results show strong agreement with the inverse design and experimental data, particularly for the parallel dipole orientation, which primarily excites multiple propagating modes dominated by TE waveguide modes. Maximum transmittances of 26.1%, 56.3% and 36.9% were achieved for output 1 (designed for 620 nm), 2 (designed for 750 nm) and 3 (designed for 870 nm), respectively, at their corresponding incidence wavelengths (see Table S1). For each wavelength, the fraction of port transmission relative to total transmission was calculated to distinguish individual port contributions, revealing over 80% of light output at the assigned ports with minimal crosstalk.

*Table S1. Parallel dipole wavelength dependency on transmittance and transmission fractions.*

| Parallel dipole wavelength, nm | Output 1, transmittance (fraction) | Output 2, transmittance (fraction) | Output 3, transmittance (fraction) |
|---|---|---|---|
| 620 | 26.1% (83.2%) | 2.3% (7.4%) | 2.9% (9.4%) |
| 750 | 3.6% (5.8%) | 56.3% (89.4%) | 3.0% (4.8%) |
| 870 | 2.4% (5.2%) | 6.1% (13.4%) | 36.9% (81.4%) |

In contrast, for the perpendicular dipole orientation excites waveguide modes dominant with TM modes. This results in reduced wavelength sorting efficiency, with maximum transmittances of 12.8%, 17.6%, and 14.1% for the target output ports (see Figure S6 and Table S2). Due to mode mismatch, since the design was optimized for the fundamental TE waveguide mode, significant crosstalk was observed across all three wavelengths.

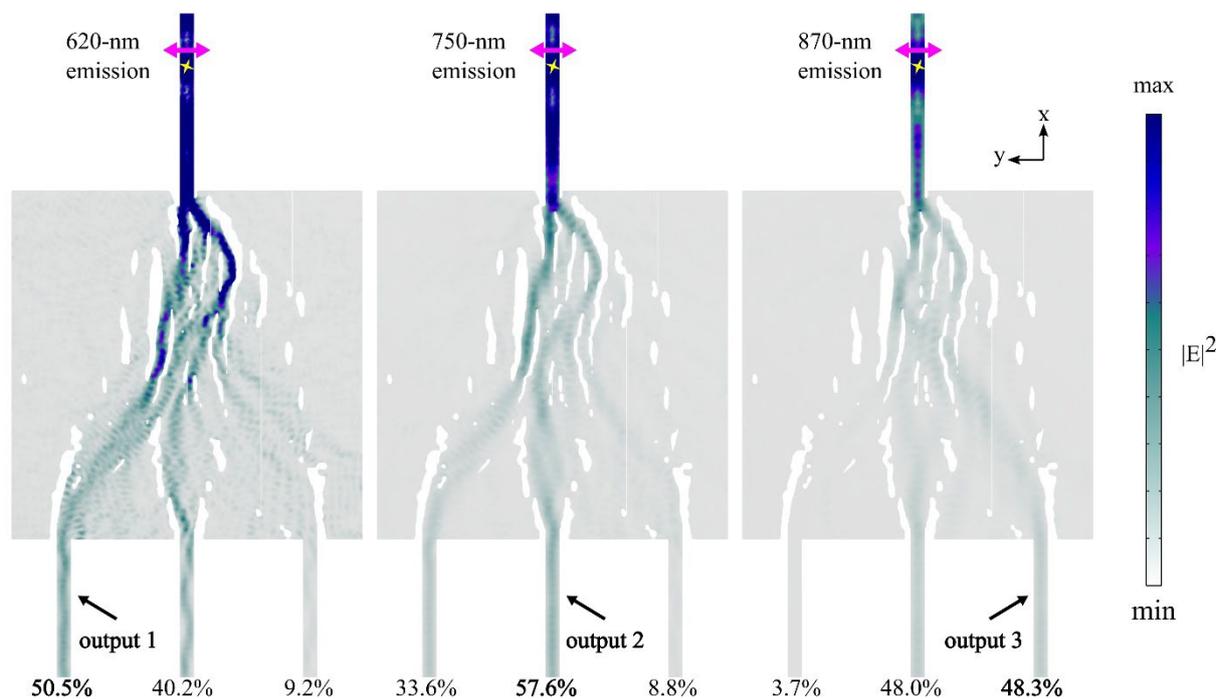

*Figure S6. Full device simulation with dipole source perpendicular to the waveguide. The numbers at each output port are transmission fractions. All plots normalized to the same value.*

Overall, the wavelength-dependent dipole emission coupling and sorting simulations confirm the reliability of our design and experimental results, enhancing our understanding of the coupling and sorting processes of excitonic emissions in the waveguide chip.

*Table S2. Perpendicular dipole wavelength dependency on transmittance and transmission fractions.*

| Perpendicular dipole wavelength, nm | Output 1, transmittance (fraction) | Output 2, transmittance (fraction) | Output 3, transmittance (fraction) |
|---|---|---|---|
| 620 | 12.8% (50.5%) | 10.2% (40.2%) | 2.3% (9.2%) |
| 750 | 10.2% (33.6%) | 17.6% (57.6%) | 2.7% (8.8%) |
| 870 | 1.1% (3.7%) | 14.0% (48.0%) | 14.1% (48.3%) |

Next, we looked at the light response to identical waveguide grating periods with numerical simulations. Figure S7 depicts guided wave extraction using grating out-couplers with 30 strips. Each column in the plot corresponds to a fixed grating period ($\Lambda$), while each row represents a consistent wavelength of incidence. Specifically, the first column features a grating period of 340 nm (designed for a wavelength of 620 nm). At the target wavelength, light is extracted homogeneously from the grating region with a small extraction angle. In contrast, other input wavelengths at this port result in anomalous scattering, typically with large deflection angles, as shown in Figures S7 b1 and S7 c1.

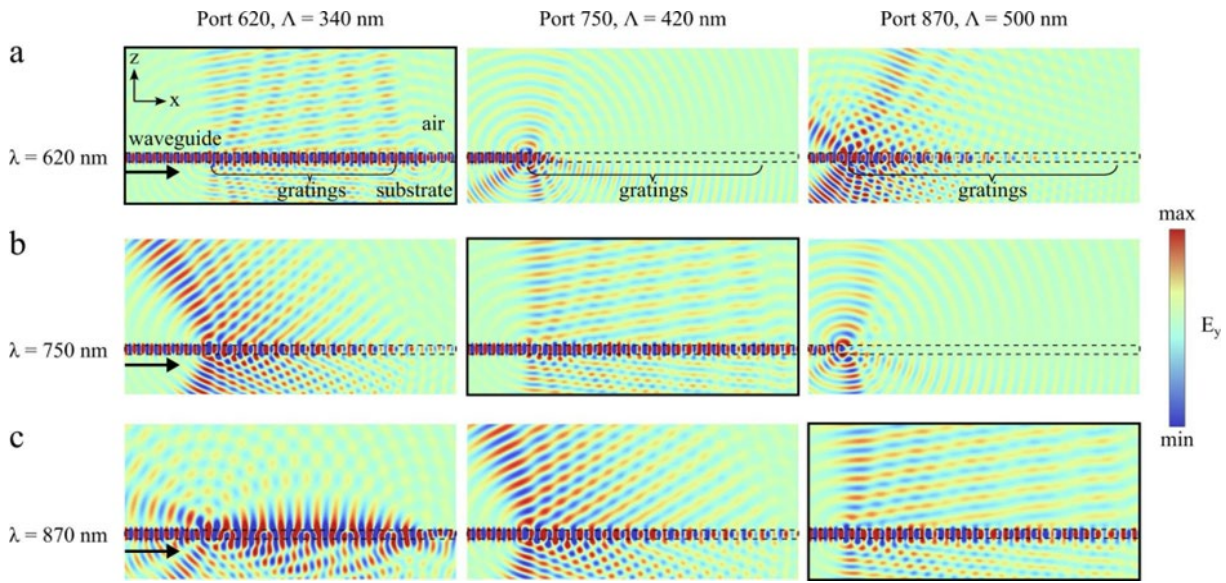

*Figure S7. Wavelength-dependent light outcoupling at three target ports. The gratings, consisting of periodic strips with a fixed filling factor of 0.65 (not shown in the figure), are designed to couple light at specific wavelengths. Panels (a), (b), and (c) depict the grating scattering responses to three guided incidences. The tailored grating periods efficiently extract the target wavelength, while other wavelengths are either scattered or reflected back into the waveguide. A slight deflection angle is incorporated to reduce reflection at the coupler region. The plots share the same colorbar.*

Similarly, for the other two ports (with grating periods of 420 nm and 500 nm), only the target wavelength, highlighted by a black frame, is efficiently extracted from the waveguide. Notably, fully etched grating couplers disrupt waveguide continuity, causing significant scattering at the coupler's beginning region (Figures a2, a3, b3, c1). To address this, shallowly etched grating couplers could be employed. As shown in Figure S8, identical grating periods yield distinct extraction angles when different emissions are incident. Alternatively, PL mapping with additional colour filters can be used to selectively evaluate the performance of the target wavelength at each port.

We remind that the demultiplexing performance, where light is sorted and separated, relies primarily on the designed demultiplexer.

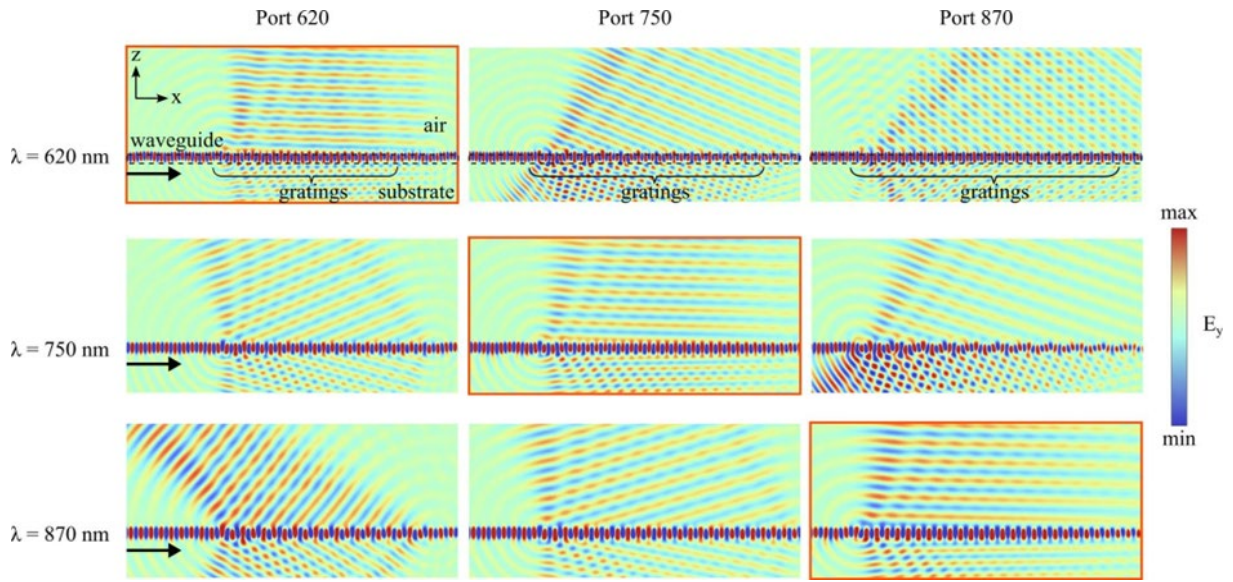

*Figure S8. Homogeneous out-coupling from shallowly etched waveguide grating couplers with a 100-nm etch depth used in the simulation. The orange-framed condition indicates the optimized grating period for each corresponding port. A small extraction angle (~5°) is intentionally introduced to minimize reflection. The plots share the same colorbar.*

**S2. Fabrication imperfections.**

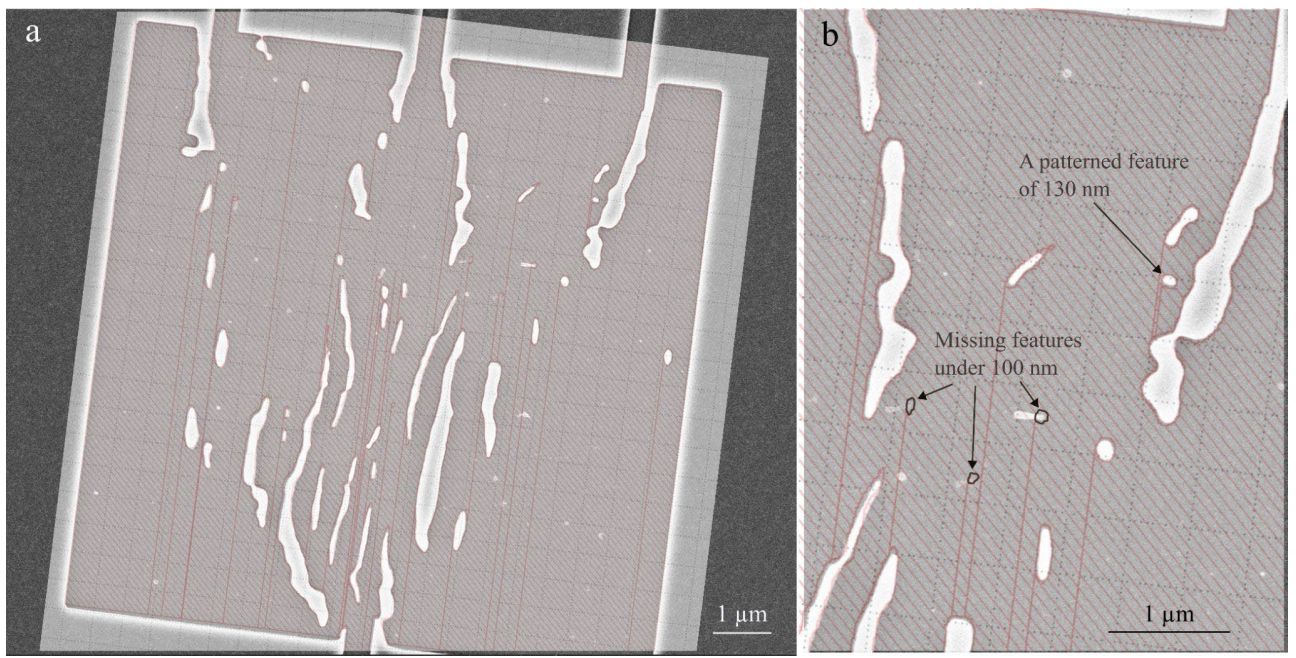

*Figure S9. Design reproducibility. (a) Scanning electron microscope (SEM) image of the device fabricated device with the overlaid design created in KLayout. (b) Enlarged area from (a) showing features smaller than 100 nm, highlighted with black lines, along with a patterned feature of 130 nm.*

The design was reproduced with high accuracy, where the smallest features of 130 nm were patterned, however the features under 100 nm were missing (see Figure S9), which might have contributed to the additional crosstalk in the experimental results. This might be due to scattered electrons during electron beam lithography that cause unintended exposure of nearby areas, making it difficult to achieve small and sharp features, as well as due to the etching process, in which small features might collapse.

## S3. Photoluminescence measurements and transmission fraction comparison with numerical simulations

After the monolayers were transferred onto the device, the presence of excitonic emission was checked. The graph in Figure S10 (a) shows the excitonic photoluminescence spectrum taken at the exciting and collecting $WS_2/WSe_2$ heterostructure from the same spot.

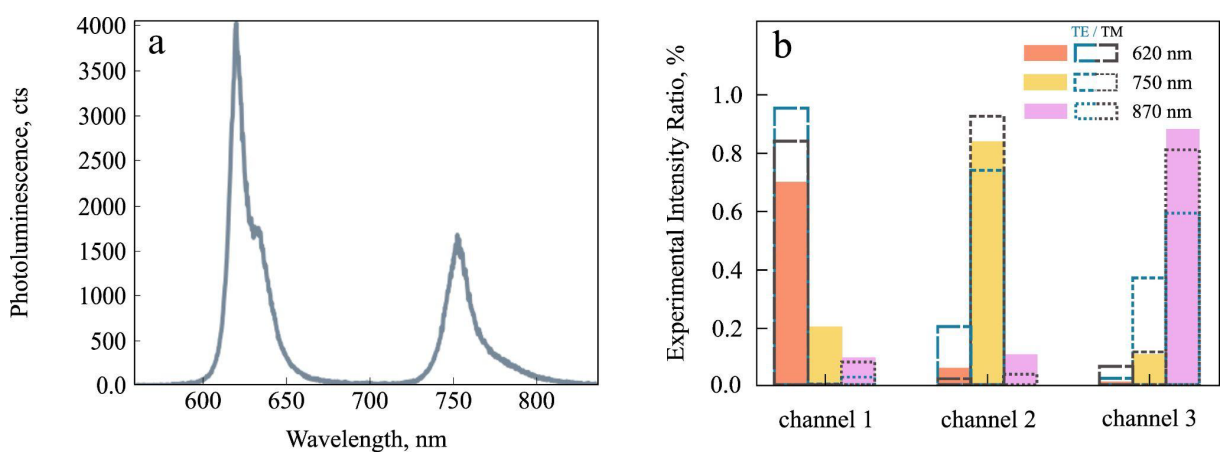

*Figure S10. (a) Photoluminescence spectra taken after the transfer of $WS_2/WSe_2$ heterostructure, excited and collected from the same spot. (b) Analysis of demultiplexer filtering extracted from experimental measurements in Figure 4 a-c and shown on the left axis, and numerical simulations in Figure S2 and shown on the right axis.*

To assess the port transmission fraction at each wavelength, a graph was plotted in Figure S10(b) using data extracted from experimental and theoretical results. The experimental data were taken from Figure 10(a-c) and plotted with the axis on the left, while the numerical simulation data were extracted from Figure S2(a) for the TE mode (blue dashed blocks) and from Figure S2(b) for the TM mode (gray dashed blocks) and plotted with the blue axis on the right. Both experimental and numerical data exhibit the same trends: most of the 620 nm emission is sorted into channel 1, the 750 nm emission into channel 2, and the 870 nm emission into channel 3, as designed.